\documentstyle[aps,epsf,preprint]{revtex}
\tightenlines
\newcommand{\infig}[2]{\begin{center}\mbox{\epsfxsize #2
\epsfbox{#1}}\end{center}}

\begin{document}


\title{Non-local two-photon correlations using interferometers physically
separated by 35 meters}

\author{W.Tittel, J.Brendel, T.Herzog, H.Zbinden and N.Gisin}

\address{University of Geneva, 
         Group of Applied Physics, 
         20,Rue de l'Ecole de Med\'ecine, 
         CH-1211 Geneva 4, Switzerland\\
         email: wolfgang.tittel@physics.unige.ch}

\date{\today}
\maketitle

\begin{abstract}
An experimental demonstration of quantum-correlations is presented. Energy and time 
entangled photons at wavelengths of 704 and 1310 nm are produced by parametric 
downconversion in KNbO3 and are sent through optical fibers into a bulk-optical (704 nm) 
and an all-fiber Michelson-interferometer (1310 nm), respectively. The two interferometers 
are located 35 
meters aside from one another. Using Faraday-mirrors in the fiber-interferometer, all 
birefringence effects in the fibers are automatically compensated. We obtained two-photon fringe 
visibilities of up to 95 \nolinebreak\% from which one can project a violation of Bell's 
inequality by 8 standard deviations. The good performance and the auto-aligning feature of 
Faraday-mirror interferometers show their potential for a future test of Bell's inequalities in order to
examine quantum-correlations over long distances.
\end{abstract}

PACS Nos. 3.65b, 42.50

\newpage

Tests of the well known Bell-inequality \cite{bell64}, which prescribes the upper limit for 
correlations between entangled particles under the assumption of local hidden variables 
theories (LHVT) \cite{bohm52}, have been made for about twenty years now. All tests are based on 
four parts: first, the source, creating entangled particles. Next, the two independent 
measuring devices, each one examining one of the particles for the correlated feature and 
ascribing a binary value to the outcome (Bernoulli-Experiment). Finally the electronics, 
putting together the results and giving thus evidence to the nonlocal correlations (see Fig. 
1). Polarization \cite{freedman72,aspect81,aspect82,switch82,ou88,kwiat95}, momentum 
\cite{rarity90}, as well as energy and time \cite{brendel92,strekalov96,tapster94} 
entangled photons have 
been used to show violations of the inequality and thus confirm the quantum-mechanical 
(QM) predictions. Even a separation of the different experimental locations by 4.3 km of 
coiled standard optical fiber corresponding to a physical distance of a few meters 
(therefore being rather time- than spacelike) did not affect the strong correlations 
\cite{tapster94}. 
However, giving up local realism has such far reaching conceptual implications that 
further experimental investigations to close remaining loopholes for LHVT are desirable. 
Moreover, the exploration of potential applications of quantum entanglement like 
quantum-cryptography \cite{ekert91}, teleportation \cite{bennett93} or more generally, 
processing of quantum information requires solid experimental grounds.

To date, all tests of the Bell-inequality are based on assumptions, which, although seeming 
reasonable, provide the means to criticize them \cite{localreal}. The experiments can be 
improved along the following lines. First, there is the so-called detection loophole: 
the detected pairs of particles could constitute a nonrepresentative sample. Second, the 
locations of the measuring devices should be spacelike separated in order to guarantee 
Einstein locality - no information on the analyzer setting on one side can propagate to the 
other side before 
both particles are detected. Hence, the settings of the two measuring devices have to be 
randomly selected after the particles have left the source. Third, the 
physical 
distance between the source and the analyzers should be increased, in order to test the QM 
predictions that the correlations do not decrease with distance. Fourth, massive particles 
should be investigated, as almost all experiments used masseless photons           
\cite{fry95}. In the 
present work, the first and the fourth point are not touched. Obviously, the two other 
points are closely related: to overcome the second one, the third one has to be mastered. 
The easiest way to achieve this, is to use optical fibers and photons at 1310 nm wavelength  
in order to keep dispersive effects and losses small.

In this letter, we present a Franson-type experiment \cite{franson89} based on energy 
and time entangled photons at center wavelengths of 704 and 1310 nm. Both photons are analyzed 
using Michelson-interferometers - the photon of the lower wavelength with a bulk-optical, 
the infrared photon with a fiber-optical one. In general, polarization mode dispersion
(PMD) has to be compensated by polarization controlling elements when working with 
fiber-optical interferometers. Using Faraday mirrors instead of regular 
mirrors in the fiber interferometer we achieve an almost perfect and automatic compensation.
Source and measuring devices are located in three different laboratories. The source is 
placed in the central lab, the 704 nm photon analyzer five meters aside in the lab next door 
and the 1310 nm photon analyzer in a third lab, located about 30 meters away in opposite 
direction down the corridor.

The Bell-inequality can be written \cite{chsh69} as a combination of four correlation 
measurements with different analyzer settings a and b (e.g. interferometric phase shift
or orientation of a polarizer)

\begin{equation}
S = \left|E (a, b) - E (a, b')\right| + E (a', b) + E (a', b') \leq 2
\label{one}
\end{equation}
and the correlation coefficient \cite {aspect82}
\begin{equation}
E(a,b) := \frac{R_{++}(a,b) - R_{+-}(a,b) - R_{-+}(a,b) + R_{--}(a,b)}
{R_{++}(a,b) + R_{+-}(a,b) + R_{-+}(a,b) + R_{--}(a,b)}
\label{two}
\end{equation}
$R_{+-}$ being e.g. the coincidence count rate between the + labeled detector at interferometer
1 and the - labeled one at interferometer 2.

In our experiment two photons entangled in energy and time are directed 
each one into an equally unbalanced interferometer. Since the path difference in each 
interferometer is much greater than the coherence length of the single photons, 
no second order interference can be observed. However, due to the entanglement, 
the possibility for the two photons to choose 
equivalent outputs can be affected by changing the phase-difference in either 
interferometer ($\delta_1$ or $\delta_2$ resp.). The quantum physical approach describes
this effect as fourth order 
interference between the probability-amplitudes corresponding to the two possibilities, 
whether the correlated photons choose both the short arms or both the long ones to 
traverse the interferometers. Due to the two remaining possibilities - the photons choose 
different arms - the visibility is limited to 50 \nolinebreak\%. However, using a fast coincidence 
technique \cite {brendel91},
the latter events can be excluded from registration thus increasing the maximum visibility
to 100 \% and leading to the coincidence probability \cite{brendel92}

\begin{equation}  
P_{ij}(\delta_1,\delta_2)=\frac{1}{8}\left(1+ijV\cos(\delta_1+\delta_2)\right),
\label{three}
\end{equation}

\noindent
where $i,j =\pm1$ and the visibility factor V describes experimental deviations from the
maximum value V = 1. Assuming detection of a representative sample of all photon pairs,
the coincidence count rate $R_{ij}$ is proportional to $P_{ij}$

\begin{equation}  
R_{ij}(\delta_1,\delta_2) \alpha \left(1+ijV\cos(\delta_1+\delta_2)\right).
\label{four}
\end{equation}
\noindent
Evaluating the correlation function (\ref{two}) with eq. (\ref{four}) leads to 

\begin{equation}
E(\delta_1,\delta_2) = V \cos(\delta_1+\delta_2).
\label{five}
\end{equation}
\noindent
Using the settings $\delta_1=\pi/4$, $\delta_1' = -\pi/4$, $\delta_2=0$, and 
$\delta_2'=\pi/2$, eq. (\ref{one}) yields 

\begin{equation}
S=V 2 \sqrt{2}\leq 2    
\label{six}
\end{equation}

\noindent
Therefore a violation of the Bell-inequality requires observing a sinusoidal correlation
function with visibility above  $1/\sqrt{2}\approx 71 \%$.

The schematic setup of the experiment is given in Fig. 2. Light from a single-line 
argon laser (140 mW at 458 nm) passes through a filter (Schott; BG 39) to separate out the 
plasma luminescence and is focused (f = 40 cm) into a KNbO3-crystal, a biaxial crystal showing 
strong nonlinear effects \cite{bbo}. Tuning the crystal temperature to 157.3 °C, 
collinear type-I phasematching for photons at 704 and 1310 nm 
is obtained for the A-cut crystal. Due to the noncritical phasematching conditions, the 
single photons exhibit rather small bandwidths of about 7 nm. Behind the crystal, the 
pump is separated out by a polarizer while the passing downconverted photons are 
launched into a standard fibercoupler. This coupler, designed to be a 3 dB (50 \%) coupler 
at 1310 nm provides uneven transmission ratios for the two output 
arms at 704 nm. 40 \% of the incoming photons are directed into the arm chosen to be the 
``visible'' one, 3.5 \% into the ``infrared'' arm and the rest is absorbed. Therefore not 
all of the pairs are split. Including additional 75 \% losses in a fiber-connection (FC) between the 
coupler and a fiber that guides only one mode at 704 nm, only 5 \% of the created photon 
pairs finally exit the source by different output fibers.

Having passed 50 m single mode fiber, the photons at the lower wavelength (704 nm) are directed 
into a Michelson-interferometer (interferometer 1) which is located 5 m aside from the 
source. A microscope-objective (MO; 40x) collimates the light at the input of the 
bulk-optical analyzer. The path-length difference, about 20 cm, is matched to the 
second, the fiber-optical interferometer (interferometer 2). This small value compared to the 
coherence length of the laser is chosen to keep phase-variations caused by temperature 
fluctuations or mechanical instabilities as small as possible. To change the 
phase-difference of interferometer 1, the mirror M1 can be moved using a translation stage which 
is driven by a computer-controlled micro-step-motor. At the output of the interferometer, 
the photons are focused into the multimode pigtail of a silicon single-photon counting 
module (EG\&G; SPCM-AQ). To ensure high visibility, a pinhole separating out only a 
single fringe is placed before the collecting lens. Since the probability for a silicon 
detector to detect an infrared photon is close to zero, all frequency filtering to block 
remaining photons at 1310 nm can be neglected.

The analyzer for the 1310 nm photons, a fiber-optical Michelson-interferometer 
(interferometer 2), is located 30 m away from the source and connected by 60 m of 
standard telecom fiber. The path-length difference is equivalent to 20 cm in air (13 cm in 
glass or 0.7 nsec time-difference) and can be changed by a computer-driven 
phase-modulator (PM) - 1.5 m of fiber wrapped around a cylindrical piezotube. Instead of 
ordinary mirrors, we use so-called Faraday-mirrors (FM), a $45^\circ$  Faraday rotator 
glued in 
front of a conventional mirror. These mirrors ensure that a photon, injected in any arbitrary 
polarization state into one of the interferometric arms will always come back exactly 
orthogonally polarized regardless any birefringence effects in the fibers. Hence, all PMD 
is automatically compensated, no polarization alignment is required any more. The 
recently reported visibility of 99.84 \% for a 23 km long interferometer used for 
quantum-cryptography \cite{muller97} demonstrates the ability of Faraday-mirrors to 
overcome any reduction 
of visibility due to PMD. The output arm of the fiber-interferometer is 
connected to a single-photon counter, a passively quenched germanium avalanche 
photodiode (Ge-APD; Siemens SRD00514H) operated in Geiger mode at 77 K. Since the 
detection-time jitter has to be smaller than the difference between traveling times along the 
two interferometric arms, the APD is biased 0.7 V above breakdown (about 23.1 V), 
leading to a jitter of 200 ps FWHM, a detectivity of 17 \% and dark-counts of about 180 
kHz. Due to the poor detectivity of the Ge-APD for visible photons, frequency filtering 
can again be ignored.

Apart from monitoring the single count rates, the signals from the single-photon detectors 
are fed into a time to pulse height converter(TPHC; EG\&G Ortec 457) in order to manifest 
the correlations between the two independent Bernoulli-experiments. We use the silicon 
detector (the detector with the lower count rate) to start the TPHC and the germanium 
detector to stop it. A window 
discriminator permits to count coincidences within a 350 ps intervall which is matched 
to the arrival time of two photons having passed equivalent paths (either short-short or
long-long). If the photons of a pair choose different paths (short-long or long-short) 
they arrive with an additional time-delay of 0.7 ns and are thus excluded from registration.
Therefore photon-pair detection is limited to the interfering processes only. 

Figure 3 shows a plot of the coincidence counts we obtained during intervals of two 
seconds as a function of the displacement of M1 in the bulk interferometer. Accidental 
coincidences, about 33 Hz, are calculated for each measured point as the product of the 
single count rates per second and the width of the coincidence-window.
They have already been subtracted. The single count rates remained constant at 250 kHz
(including 60 Hz dark) for the silicon and 380 kHz for the germanium detector 
(including 180 kHz dark). Fitting the experimental data by a sinusoidal curve, we 
find a visibility of 95.7 \%$\pm$3.15 \%. Comparable results for a change of the path-length
difference 
in the fiber-interferometer have been obtained, too. Our experimental setup allows to observe only
one output of each analyzer. Hence, only one of the four coincidence functions in eq. (4),
requested to calculate the correlation function (eq. (5)), can be detected. From symmetry considerations,
it is reasonable to assume that the measurement of the 3 other coincidence functions would show the same 
visibility. Assuming this, we can deduce a violation of the Bell-inequality by 8 standard 
deviations \cite{visibility}.

Apart from problems with thermal stability, deviation from the maximum 
theoretical value for visibility of 100 \% is mainly caused by the mismatch of the spatial modes in the bulk 
interferometer. Due to the birefringence-compensating effect of the Faraday-mirrors and 
chromatic dispersion close to zero, originating from the small single-photon bandwidth, 
contributions by the fiber interferometer are negligible

In conclusion, strong quantum correlations over 35 meters have been observed. Under the above
made assumptions of symmetry, the obtained fringe visibilities yield significant violations
of the Bell-inequality. Using Faraday-mirrors in the fiber interferometer, all 
polarization control can be disregarded. This is a major 
advantage compared to already existing schemes using optical fibers. Employing two 2-output 
auto-aligning interferometers with improved thermal stability and creating both 
entangled photons at telecom wavelength of 1310 nm will implement a long distance 
Bell-experiment with tens of kilometers of spatial separation between the source and each of 
the measuring devices. This in turn will open the possibility to test quantum-mechanics 
with truly random settings obeying Einstein locality ore more general, will lay grounds for 
further processing of quantum information.

We would like to thank the quantum-optical group in Frankfurt/Main (Germany) directed 
by Prof. W. Martienssen for many helpful discussions and for placing the nonlinear
crystal to our disposal. This work is supported by the 
Swiss Priority Program ``Optique''. J. Brendel and T. Herzog are financially 
supported by the TMR network on the Physics of Quantum Information.


\begin{figure}
\infig{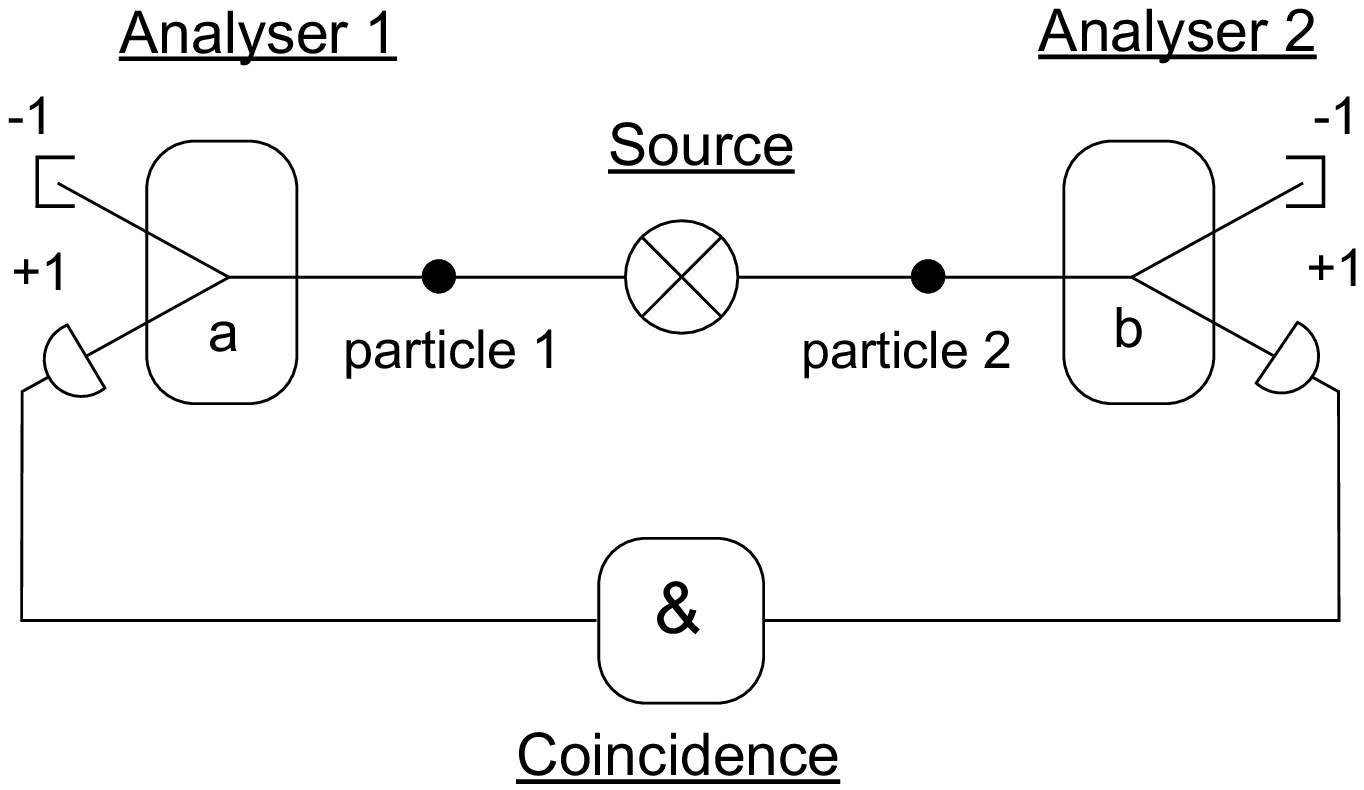}{0.85\columnwidth}

\caption{Principle setup to test the Bell-inequality. The entangled particles, created by a two-
particle source, are separated and each one sent to an analyzer. Each measuring device 
performs a Bernoulli-experiment with binary valued output depending on the correlated 
feature. The results - in our case only the +1 labeled  events - are put together, giving thus 
evidence to the nonlocal correlations.} 
\label{fig1} 
\end{figure}

\begin{figure}
\infig{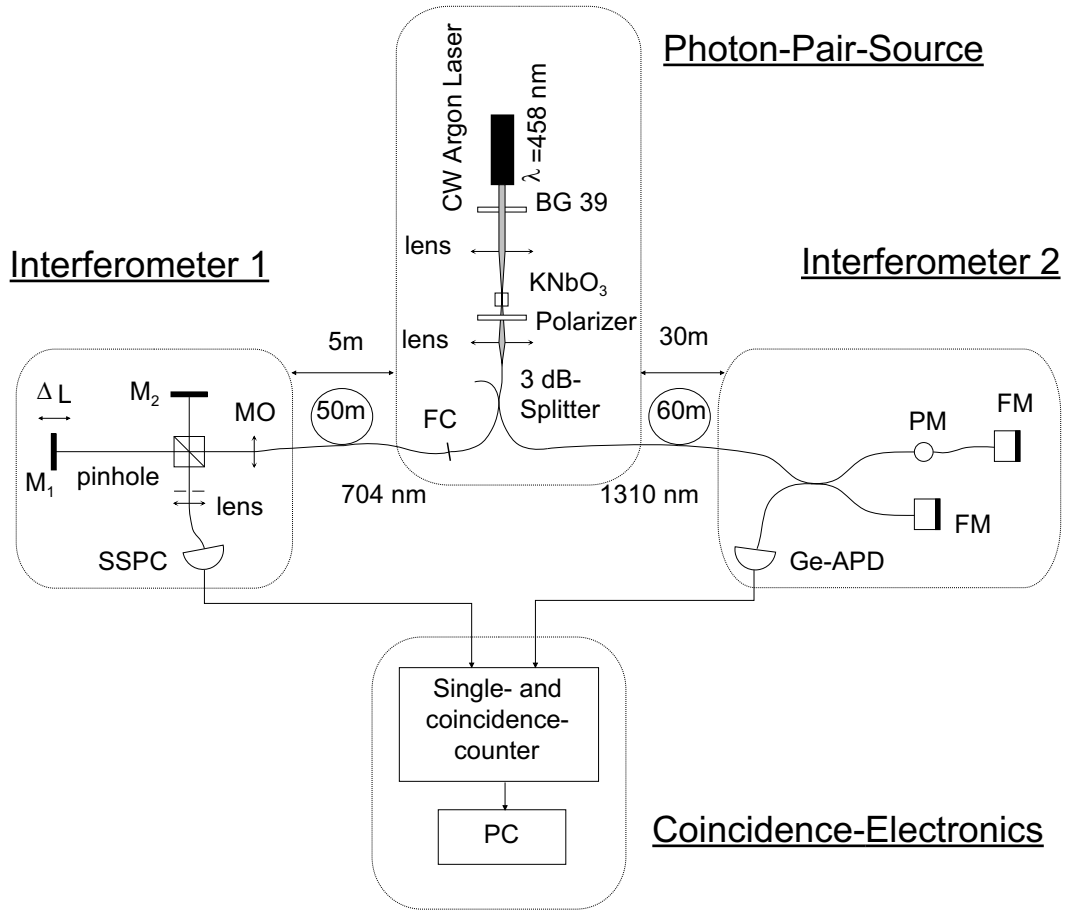}{0.85\columnwidth} 
\caption{Experimental setup of the Franson-type test, see text for detailed 
description.\hspace*{1cm}}
\label{fig2}
\end{figure}

\begin{figure}
\infig{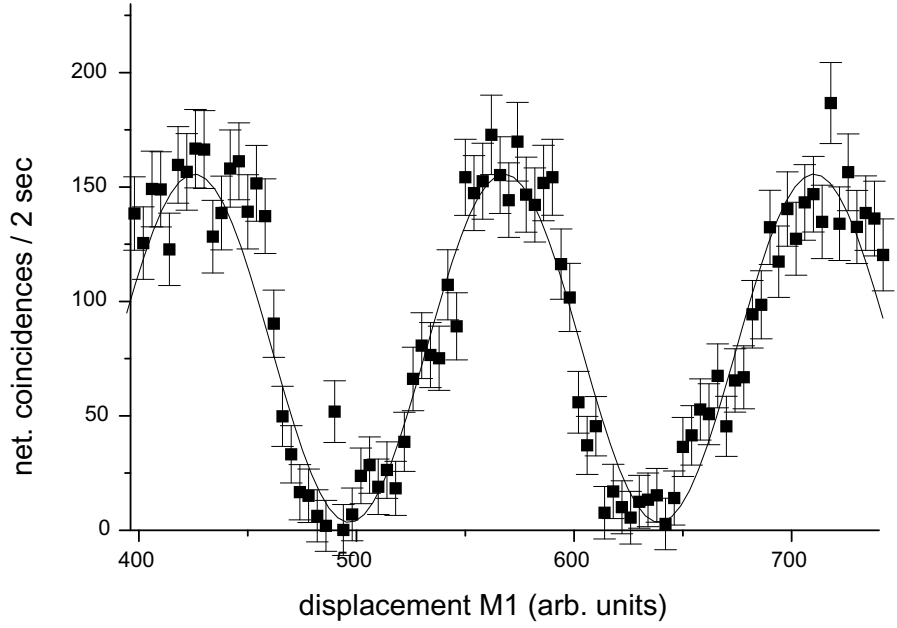}{0.85\columnwidth} 
\caption{Net coincidence counts per 2 sec. as a function of path-length difference in the 
bulk-interferometer (displacement M1). Best fit with a sinusoidal function yields a visibility of 
95.7 \%$\pm$3.15 \%.}
\label{fig3}
\end{figure}


\begin{references}
\bibitem{bbo}The corresponding coefficient for the used A-cut $d_{32}$ = -18.3 pm/V 
        is about nine times the highest coefficient in BBO.
\bibitem{visibility}To determine the visibility corresponding to the measured data, we apply a 
                   least squares fit ($\chi^2$-test), assuming the sinusoidal function (\ref{four}) 
                   with V as 
                   the free parameter. The reliability of the best fit can be found by looking at the 
                   tolerated variation of the fitted parameters in order to keep the $\chi^2$ value 
                   smaller than a certain limit. This threshold, depending on the number of 
                   measured data and the desired confidence level (as common in statistics, we 
                   choose 0.95) can be found in tables. Thus, the probability that our data are 
                   corresponding to a sinusoidal function with a visibility of $81.6\pm1.1\%$ is 0.95.
\end{references}
\end{document}